\documentclass[pt12]{JHEP3} 

\JHEPspecialurl{http://jhep.sissa.it/JOURNAL/JHEP3.tar.gz}

\usepackage{epsfig,multicol}

\usepackage{amsmath, amssymb,graphicx, epic, eepic}

\newcommand{\RR}{\mathbb{R}}
\newcommand{\CC}{\mathbb{C}}
\newcommand{\ZZ}{\mathbb{Z}}

 \newcommand{\N}{{\cal N}}

\title{Mesons in gauge/gravity dual with large number of fundamental fields}

\author{Johanna Erdmenger and Ingo Kirsch\\
        Institut f\"ur Physik,\\ Humboldt-Universit\"at zu Berlin,\\
Newtonstra{\ss}e 15,\\ D-12489 Berlin,\\ Germany\\
        E-mail: \email{jke@physik.hu-berlin.de},
                \email{ik@physik.hu-berlin.de}}

\received{***, 2004}            
\revised{***, 2004}
\accepted{***, 2004}            

\preprint{\hepth{0408113}\\ HU-EP-04/42}                                 

\abstract{In view of extending gauge/gravity dualities with flavour beyond the
  probe approximation, we establish the gravity dual description of mesons 
  for  a
  three-dimensional super Yang-Mills theory with fundamental matter. For this
  purpose we consider the fully backreacted D2/D6 brane solution of Cherkis and
  Hashimoto in an approximation due to Pelc and Siebelink.  The low-energy
  field theory is the IR fixed point theory of three-dimensional $\N=4$
  $SU(N_c)$ super Yang-Mills with $N_f$ fundamental fields, which we consider
  in a large $N_c$ and $N_f$ limit with $N_f/N_c$ finite and 
  fixed. We discuss the
  dictionary between meson-like operators and supergravity fluctuations in the
  corresponding near-horizon geometry. In particular, we find that the mesons
  are dual to the low-energy limit of closed string states. In analogy to
  computations of glueball mass spectra, we calculate the mass of the
  lowest-lying meson and find that it depends linearly on the quark mass.}


\dedicated{}

\begin{document}

\section{Introduction}

The AdS/CFT correspondence may be generalized such as to describe
degrees of freedom in the fundamental representation of the gauge
group. This was first suggested in \cite{Karch} where a D7 brane probe
is embedded in the AdS/CFT D3 brane background. Fundamental degrees of
freedom are given by open strings stretching between the D3 branes and
the D7 brane probe. These are dual to open strings on the D7 probe
wrapping $AdS_5 \times S^3$, as given by the Dirac-Born-Infeld action
for the D7 brane probe in the low energy limit.  The associated field
theory is a $\N=2$ supersymmetric Yang-Mills theory in $d=4$ with
$N_7$ fundamental hypermultiplets, where $N_7$ is the number of D7
branes.

The meson spectrum for the $\N=2$ supersymmetric theory of
\cite{Karch} is calculated in~\cite{Kruczenski}. The meson masses are
shown to scale linearly with the quark mass as required by
supersymmetry. An essential ingredient in this calculation is to
consider the limit $N_7 \ll N_3$ with $N_7$ fixed, such that the
backreaction of the D7 probe on the geometry can be neglected.  A
similar calculation of the meson spectrum for the $\N=1$
supersymmetric Klebanov-Strassler background with an embedded D7 brane
probe is performed in \cite{Sonnenschein}.

In \cite{Babington}, the D7 brane probe is embedded into the
non-supersymmetric Constable-Myers background \cite{CM}.\footnote{See also
  \cite{Evans} for further results on the probe configuration in this
  background.}  In this way a gravity
dual description of spontaneous chiral symmetry breaking is found, including a
Goldstone boson which was noted to have properties of the $\eta'$ meson: 
The $\eta'$ is a Goldstone boson of $U(1)$ only in the limit 
$N \rightarrow \infty$. This is just the limit in which
the supergravity approximation is valid. 
Moreover in \cite{Babington} it is shown that the pseudo-Goldstone
boson mass scales with the square root of the quark mass in agreement with
field theory expectations. In~\cite{Kruczenski:2003uq}, a similar computation
is performed within type IIA supergravity for the D4 brane black hole metric
of \cite{Wittenblackhole} with a D6 brane probe.  This background metric has
the advantage of being regular in the interior (ie.~in the infrared of the
field theory).  Spontaneous chiral symmetry breaking occurs in this background
as well.  Moreover, a holographic version of the Vafa-Witten theorem is
derived in this brane set-up~\cite{Kruczenski:2003uq}.  - A D5 brane probe in
the Maldacena-Nu$\rm \tilde n$ez background is investigated in \cite{Ramallo}
(see also \cite{Wang:2003yc}).

All of these calculations are performed in the probe approximation,
neglecting the backreaction of the probe brane. In view of further
progress it is necessary to go beyond both the probe limit and the
supergravity approximation. A step in this direction was taken in
\cite {KMB}, where $1/N$ corrections to the supergravity limit were
considered in order to derive the Witten-Veneziano formula for the
$\eta'$ meson in the D4/D6 probe configuration.  Further related
results are also found in \cite{Armoni}. In \cite{Vaman} a probe
D7 brane is embedded in the backreacted D3/D7 solution, and the spectrum
of scalar fluctuations is computed to first order in $N_f$.

In this paper we take a step towards describing a large number of
fundamentals by computing the meson spectrum in a full supergravity
metric for intersecting branes {without} embedding a probe brane. To
this effect, we consider the type IIA D2/D6 brane system for which the
supergravity metric is known \cite{Cherkis} for arbitrary $N_2$ and
$N_6$, the number of D2 and D6 branes, respectively.  The low energy
dynamics of this system is described by a 2+1 dimensional $\N=4$
$SU(N_2)$ super Yang-Mills theory which is coupled to an adjoint and
$N_6$ fundamental hypermultiplets. The theory, whose gauge coupling
is dimensionful, flows to a superconformal
fixed point in the~IR.

The D2/D6 configuration is T-dual to both the D3/D5 configuration
considered in \cite{DeWolfe, Erdmenger} and to the D3/D7 configuration
of \cite{Karch}. Recently, the backreaction of D7 branes in the $AdS_5
\times S^5$ and $AdS_5 \times T^{1,1}$ geometries were studied in
\cite{Vaman}. However, as compared to the backreacted D3/D7 configuration,
which leads to a 3+1 dimensional theory with flavour, the D2/D6 system
has the essential technical advantage that the backreaction of the D6
branes does not lead to a deficit angle.  The field theory corresponding
to D3/D7 is not conformal if $N_f/N_c$ is kept finite and fixed.
In contrast, the three-dimensional theory corresponding to D2/D6, which 
is superrenormalizable, flows to a non-trivial IR fixed point 
even for $N_f/N_c$ finite.

The solution for the fully localized D2/D6 brane intersection was
given by Cherkis and Hashimoto in \cite{Cherkis}. We consider this
solution in the limit $1 \ll N_6 \ll N_2$ with $N_6/N_2$ fixed such
that the backreaction of the D6 branes must be taken into
account. This geometry has the important feature that the fluctuations
of the D6 branes induce fluctuations of the supergravity background.
The corresponding near-horizon geometry describes the full
renormalization group flow of the 2+1 dimensional super Yang-Mills
theory from the UV to the interacting fixed point in the IR. For the
purpose of computing the meson spectrum, it is enough to focus on the
supergravity description of the IR fixed point theory which is given
by the metric of Pelc and \mbox{Siebelink}
\cite{Pelc}. This metric approximates the region close to the center
of the Cherkis-Hashimoto background corresponding to the IR regime of
the field theory.

Near the boundary, the Pelc-Siebelink metric reduces to $AdS_4 \times
S^7/\ZZ_{N_6}$. However, the interior of this geometry is deformed due to a
non-vanishing quark mass. The corresponding Kaluza-Klein spectrum has been
worked out by Entin and Gomis \cite{Gomis}. Moreover, these authors relate
some of the supergravity excitations to chiral operators in the
2+1-dimensional theory.  Chiral operators belong to the short representations
of the superconformal algebra and as such they are protected against quantum
corrections. For this reason these operators remain unrenormalized in the IR
fixed point theory, where they are dual to the supergravity fluctuations on
$AdS_4 \times S^7/\ZZ_{N_6}$. We extend this analysis to meson-like operators
involving components of the fundamental superfields in view of computing the
meson spectrum.

Using mirror symmetry we argue that the field theory operators
involving fundamentals are dual to supergravity modes, ie.~to the low
energy limit of closed strings. This is in contrast to the probe
approximation, where the meson-like operators are dual to open strings
on the probe brane.  Another meson/closed string duality has been
proposed in \cite{Armoni}, in which D3-branes are considered on an
orbifold of the type $\CC^3/(\ZZ_3 \times \ZZ_3)$. There the
lightest pseudo-scalar meson is conjectured to be dual to a particular
twisted RR field which is generated by closed strings.

The conjectured meson/closed string duality allows us to compute the meson
spectrum by a supergravity computation in the Pelc-Siebelink background.
Similarly as in calculations of the glueball mass spectrum \cite{Csaki,
  deMelloKoch:1998qs, Zyskin:1998tg, Ooguri:1998hq}, we use a plane wave
ansatz for the fluctuations of the scalar field dual to the mesons. We
numerically solve the resulting equation by using standard shooting
techniques. The quark mass enters the Pelc-Siebelink metric as a free
parameter. Varying this parameter, we determine the mass of the lowest-lying
meson in the spectrum and its dependence
of the quark mass. We find a linear behaviour on the quark mass as expected
for supersymmetric field theories.

The outline of the paper is as follows. In section 2 we present the
supergravity setup for the D2/D6 brane configuration, and discuss the
$2+1$-dimensional $\N=4$ supersymmetric gauge theory associated with this
background.  In section 3 we provide the gauge/gravity dictionary for this
set-up.  The main feature is that the field theory operators involving
fundamental fields are dual to closed strings in the Pelc-Siebelink geometry.
These closed strings are generated by the fluctuations of the D6 branes. We
also calculate the lowest-lying meson mass for the gauge theory from
supergravity and find it to scale linearly with the quark mass, in agreement
with field theory expectations.

\medskip

\section{Holography of the D2/D6 brane intersection}

\subsection{Configuration}

We consider $N_2$ D2-branes with world-volume coordinates along $x^0,
x^1, x^2$ parallel to $N_6$ D6-branes along $x^0,...,x^6$. This brane
intersection preserves 8 supercharges and is T-dual to the orthogonal
D3/D$p$ brane intersections ($p \in \{3,5,7 \}$) whose holography
is studied in \cite{Karch, DeWolfe, Constable:2002xt,  Kirsch:2004km}.

\begin{figure}
\begin{center}
\includegraphics[scale=.8]{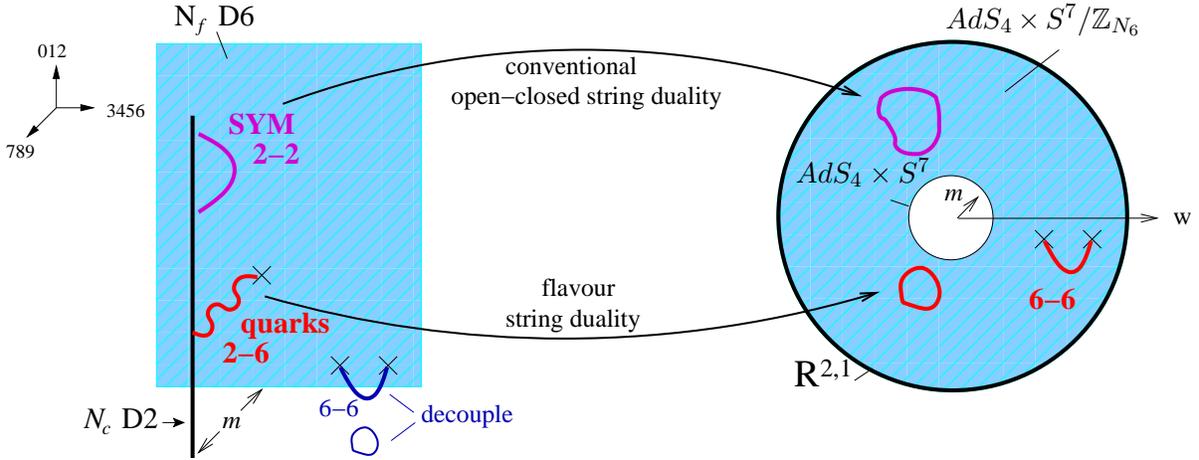}
\end{center}
\caption{Two-fold open-closed string duality: a) standard duality between
adjoint operators (2-2 strings) and a class of closed strings. b)
flavour string duality between meson operators (2-6 strings) and
another class of closed strings. This will be discussed in detail in
Sec.~3.1.}  \label{fig1}
\vspace{-8.4cm} \hspace{12.4cm} $AdS_4 \times S^7/\ZZ_{N_6}$ \vspace{7.9cm}

\vspace{-6.9cm} \hspace{10.4cm} $AdS_4 \times S^7$ \vspace{6.4cm}
\end{figure}

There are several kinds of strings in the D2/D6 set-up. First, there
are open strings ending on the D2-branes (2-2 strings) and open
strings stretching between the D2- and the D6-branes (\mbox{2-6} and
6-2 strings). Closed strings decouple upon taking the usual $l_s
\rightarrow 0$ limit. We also send $g_s \rightarrow 0$ such that the
2+1-dimensional gauge coupling
\begin{align}
  g^2_{\rm YM2}= g_s l_s^{-1}  
\end{align}
remains fixed. In order to keep the 't~Hooft
coupling $\lambda_2$ constant, we take the limit
\begin{align} \label{limit}
 N_2, N_6 \rightarrow \infty\,, \quad 
 g^2_{\rm YM2} \rightarrow 0\,, \quad
\lambda_2 \equiv g_{\rm YM2}^2 N_2 ={\rm const} \,,\quad
 \nu \equiv N_6/N_2={\rm const.}
\end{align}
The 6+1-dimensional 't~Hooft coupling
\begin{align}
  \lambda_6 \equiv g^2_{\rm YM6} N_6 = (2 \pi)^4 g_s l_s^3 N_6 
            = (2\pi l_s)^4 g^2_{\rm YM2} N_6 
            = (2\pi l_s)^4 \lambda_2 \nu   
\end{align}
vanishes in this limit implying the decoupling of open 6-6
strings. The massless open string degrees of freedom of the D2/D6
intersection correspond to a $\N=4, d=3$ super Yang-Mills multiplet
coupled to $N_6$ hypermultiplets in the fundamental representation and to
one hypermultiplet in the adjoint representation of the gauge group.
The D2/D6 intersection and the decoupling of the strings is
shown on the left hand side of Fig.~\ref{fig1}. 

The full supergravity solution for the D2/D6 intersection including
the backreaction of the D6-branes has been found by Cherkis and
Hashimoto \cite{Cherkis}. Note that in the limit~(\ref{limit}), the
backreaction of the D6-branes on the geometry is not negligible as can
be seen by comparing the tension of both stacks of branes,
\begin{align}
T_{\rm D6} = \frac{\nu}{(2\pi l_s)^{4}} \,T_{\rm D2} \,,
\end{align}
which is very large at low energies ($l_s \rightarrow 0$) if $\nu$ is
fixed. 

The near-horizon limit of the Cherkis-Hashimoto metric can be obtained
upon taking the coupling $g^2_{\rm YM2} \rightarrow \infty$. The
D2-branes are then completely resolved by their near-horizon geometry,
whereas the D6-branes remain in the background. A singularity in the
metric signals the existence of open strings ending on the D6-branes
\cite{Pelc}. At strong coupling, we thus have closed string modes arising  
from both the D2 and the D6-branes and open strings attached to
the D6-branes. Fig.~\ref{fig1} shows the relevant strings in the
near-horizon geometry.\footnote{To be more precise, Fig.~\ref{fig1}
shows the strings in the corresponding {\em ten-dimensional}
near-horizon geometry which follows by dimensional reduction.}

A summary of the string dualities in the D2/D6 brane configuration is as
follows.\footnote{A detailed discussion of the fluctuation-operator map is
  given in Sec.~\ref{secflucmap} below.} In the limit given by (\ref{limit}),
we have two kinds of open strings on the gauge theory side of the
correspondence: open strings stretching between the D2 and D6-branes (2-6
strings), and open strings ending on the D2-branes (2-2 strings). In the
low-energy theory these strings give rise to composite operators of purely
adjoint fields and to meson-like operators involving fundamental matter
fields. Each of these operators maps unambiguously to a supergravity
fluctuation generated by closed strings. This splits the closed string modes
into two classes: Those which are dual to purely adjoint operators and those
dual to meson operators. As shown in Fig.~\ref{fig1}, we conjecture the
AdS/CFT duality to act `twice' here: First, we have the standard AdS/CFT
duality between open 2-2 strings and a class of closed strings in type IIA
string theory. Secondly, we have a further duality between open 2-6 strings
and another class of closed strings. In the low-energy limit, the two
classes of closed strings lead to different supergravity multiplets. This
flavour open-closed string duality will be relevant for the computation of the
meson spectrum.

\subsection{Field theory}

In the following we discuss the world-volume theory of the D2/D6
intersection, a $\N=4$ $SU(N_c)$ super Yang-Mills
theory ($N_c \equiv N_2$) which is coupled to an adjoint and $N_f
\equiv N_6$ fundamental hypermultiplets. The action can be
conveniently written in $\N=2$, $d=3$ superspace formalism and is
derived in appendix~\ref{appA}. The theory flows to a superconformal
fixed point in the IR \cite{Ferrara:1998vf}. It belongs to a class of
three-dimensional super Yang-Mills theories studied in a number of
papers \cite{deBoer, Aharony, Seiberg, Intriligator:1996ex,
Kapustin:1999ha, Kitao:1998mf, Kitao:1999uj}.

First, let us consider the system in the absence of D6-branes.  The low-energy
effective action on the D2-branes is 2+1-dimensional $\N=8$ SYM theory
with gauge group $SU(N_c)$ which is generated by the dynamics of 2-2
strings.  The theory has a dimensionful gauge coupling and is not
conformal in the UV. The gauge coupling is determined by the type IIA
string coupling which blows up in the IR. At strong coupling, it is
more appropriate to consider this theory as the world-volume theory of
M2-branes instead of D2-branes whose near-horizon geometry is $AdS_4
\times S^7$.  This agrees with the behaviour of the field theory which
flows to a superconformal fixed point in the IR \cite{Ferrara:1998vf}.

By adding $N_6$ D6-branes, we break half of the supersymmetry such that the
theory on the intersection becomes a 2+1-dimensional $\N=4$ gauge theory.
Its field content is as follows. The ${\cal N}=8$ vector multiplet
splits into a $\N=4$ vector and a $\N=4$ adjoint hypermultiplet or,
equivalently, into three $\N=2$ chiral superfields $\Phi_1, \Phi_2$,
$\Phi_3$, and a $\N=2$ vector multiplet $V$.  In addition we have $N_f
\equiv N_6$ $\N=4$ fundamental hypermultiplets (``quarks'') generated
by the low-energy dynamics of 2-6 strings. These hypermultiplets
consist of $\N=2$ chiral superfields $Q_f$ and $\tilde Q_f$ with the
flavour index running from 1 to $N_f$. The theory has thus a global
$U(N_f)$ flavour symmetry. The decoupling of the 6-6 strings guarantees
that the the $U(N_f)$ symmetry on the world-volume of the D6-branes
acts as a global symmetry rather than a gauge symmetry. Note also that
due to the indistinguishability of left- and right-handed fermions,
there is no chiral symmetry $U(N_f) \times U(N_f)$ in three
dimensions. 

$\N=4$ gauge theories in three dimensions have an $OSp(4/4)$
supergroup.  Its even part is the product of the conformal group
$SO(3,2)$ and an $SU(2)_V \times SU(2)_H$ R-symmetry group.  The
$SU(2)_V$ symmetry is the enhanced $U(1)_R$ symmetry of $\N=2, d=4$
supersymmetry upon reduction to three dimensions and is realized in
the dual string theory as rotations in the 789
directions.\footnote{For a detailed discussion on the R-symmetry
enhancment see \cite{Seiberg}.}  As a subgroup of an anomaly-free
group, the $U(1) \subset SU(2)_V$ is non-anomalous, unlike $U(1)_R$ in
four dimensions.  The $SO(4) \approx SU(2)_H \times SU(2)_\Phi$
isometry in the 3456 directions consists of a global $SU(2)_\Phi$
symmetry rotating $\Phi_1$ and $\Phi_2$ and the $SU(2)_H$ part of the
$\N=4$ R-symmetry.

\begin{table}[ht]
\begin{center}
\begin{tabular}{cclccccc}
\N=4 & \N=2 &  components  & spin & $j_\Phi$& $(j_H,j_V)$ 
&$U(N_f)$&$\Delta$  \\
\hline & & $\sigma, \phi_3$ & $0$&
$0$ &$(0,1)$ &$1$ &${1}$ \\
vector& $\Phi_3, \Sigma$ &  $\lambda_{3}, \bar\lambda$ & $\frac{1}{2}$&
$0$ &$ (\frac{1}{2},\frac{1}{2})$&$1$& $\frac{3}{2}$ \\
& & $A_i$  & $1$ & $0$ &$(0,0)$&$1$&$1$\\
\hline
  adj.& & $\phi_1$, $\phi_2$ & $0$ & $\frac{1}{2}$ &$(\frac{1}{2},0)$ &$1$&
  $\frac{1}{2}$ \\ hyper& $\Phi_1, \Phi_2$ & $\lambda_{1},
  \bar\lambda_{2}$ & $\frac{1}{2}$ & $\frac{1}{2}$ & $(0,\frac{1}{2})$&$1$& $1$
\\
\hline
 fund.& & $q_f$, $ {\tilde q_f}$  & $0$& $0$ & $(\frac{1}{2},0)$ &$N_f$&
$\frac{1}{2}$\\ 
hyper& $Q_f, \tilde Q_f$ & $\psi_f, \tilde \psi_f$ & $\frac{1}{2}$ & $0$ &
$(0,\frac{1}{2})$ &$N_f$& ${1}$ 
\end{tabular}
\caption{Field content of the D2/D6 intersection.}\label{rsym}
\end{center}
\end{table}

Table~\ref{rsym} summarizes the R-charges and engineering dimensions of the
fields of the D2/D6 intersection. The dimension $\Delta$ and the R-symmetry
quantum numbers $j_V$ and $j_H$ of the scalars saturate the inequality $\Delta
\geq j_V + j_H$ \cite{Seiberg2}. For this reason, the conformal dimensions of
scalars and spinors in the vector multiplet differ from those of scalars and
spinors in the hypermultiplets.  In components the theory contains seven
adjoint scalars $\phi_{1,2,3}$, $\sigma$ and four adjoint spinors
$\lambda_{1,2,3}$, $\lambda$ (gaugino). There is also the gauge field $A_i$
which can be explicitly dualized to a scalar only in the abelian case. The
components of the fundamental hypermultiplets are the scalars $q_f$, $\tilde
q_f$ and the spinors $\psi_f$, $\tilde \psi_f$.

\subsubsection{Mass terms and explicit flavour symmetry breaking on
the Higgs branch}

Adding mass terms for the fundamental fields corresponds to separating
the D2-branes from the D6-branes in the transverse directions $x^7,
x^8, x^9$. These directions are para\-metrized by the expectation values
of the complex scalar $\phi_3$ and the real scalar $\sigma$. These are
the lowest components of the scalar and vector multiplets $\Phi_3$ and $V$,
respectively. The 2-6 strings give rise to quarks with a
(tree-level) mass $m=\sqrt{\langle\sigma\rangle ^2 +
\vert \langle \phi_3 \rangle \vert^2}$ via the interaction terms $\langle\sigma
\rangle \bar \psi \psi$ and $\langle \phi_3 \rangle
\tilde \psi \psi$. In contrast to the four-dimensional field theory 
corresponding to the D3/D7 system, the $U(1)$ group inside the global
symmetry group $U(N_f)$ is a symmetry for any choice of quark
masses~\cite{Ferretti}.\footnote{In the four-dimensional theory
associated with the D3/D7 system
\cite{Karch} massive quarks explicitly break the chiral $U(1)$
symmetry.} This can also be seen in the D-brane set-up, where the
$U(1)$ group is a subgroup of $SU(2)_V$ which rotates the directions
$x^7, x^8, x^9$. A vacuum expectation value for the vector
$X_V=(\sigma, {\rm Re}\, \phi_3, {\rm Im}\,
\phi_3)$ is invariant under rotations on a two-plane
transverse to the vector $X_V$. This corresponds to the $U(1)$ subgroup.

In a three-dimensional field theory in which all quarks have equal
masses, parity is broken. There is however a way to preserve 
parity: The D6-branes have to be devided into two stacks of $N_6/2$ 
D6-branes, where one stack is located a
distance $-m$, the other one a distance $+m$ away from the D2-branes.
Then the quarks come in pairs with equal magnitude but opposite sign
leading to a parity-preserving field theory \cite{Ferretti,
Aharony}.\footnote{Such a brane set-up lifts to a stack of $N_2$
M2-branes at a double Taub-NUT space (DTN) which is a multi-Taub-NUT
space with two centers corresponding to two stacks of D6-branes. The
core of the DTN geometry (corresponding to the IR regime in the field
theory) is approximated by the Eguchi-Hanson metric. A numerical
supergravity solution corresponding to M2-branes in an Eguchi-Hanson
space has recently been found in \cite{Clarkson}.}

In theories with an even number of flavours, such mass terms break
the flavour group $U(N_f)$ down to $U(N_f/2) \times
U(N_f/2)$ explicitly.  The Higgs branch of the field theory considered here is
parametrized by $(N_f/2)^2$ mesons $M^i_j=Q^i \tilde Q_j$, where
$Q_i$, $\tilde Q_j$ are the fundamental chiral multiplets with indices
$i,j=1,...,N_f/2$ \cite{Aharony}. These mesons are elements of the
coset space $U(N_f)/(U(N_f/2) \times U(N_f/2))$. More on the Higgs as well
as on the Coulomb branches can be found in \cite{Aharony}.

\subsection{D2/D6 background and its near-horizon limit}

In the previous section we have discussed the D2/D6 brane intersection
in terms of its world-volume field theory. We now study the dual
supergravity description and review some properties of the
corresponding near-horizon geometry.

Cherkis and Hashimoto derive the D2/D6 brane solution \cite{Cherkis}
by lifting the D2/D6 system to M-theory, where the D2-branes become
M2-branes and the D6-branes turn into a Taub-NUT space. After solving
a harmonic equation, the geometry is reduced back to ten
dimensions. This approach guarantees the preservation of
8~supercharges.

The Cherkis-Hashimoto background encaptures the full renormalization
group flow of the 2+1 dimensional field theory to the fixed point in
the IR. In the following we focus on the simpler supergravity solution
found earlier by Pelc and Siebelink \cite{Pelc}. This solution is dual
only to the superconformal IR fixed point theory.

Pelc and Siebelink exploit the fact that close to the center of the
Taub-NUT space, the geometry is approximated by the orbifold
$\CC^2/\ZZ_{N_6}$. The D2/D6 system lifts to M2-branes at
$\CC^2/\ZZ_{N_6}$, as opposed to M2-branes at a Taub-NUT space (as in
the Cherkis-Hashimoto background).  The corresponding
eleven-dimensional near-horizon metric is given by
\cite{Pelc}
\begin{align} \label{11dPelc}
ds^2 = \frac{w^2}{16 (LF)^{2/3}} dx^2_{||} + (LF)^{1/3} 
\left( \frac{dw^2}{w^2} + d \tilde \Omega_7^2 \right) \,,
\end{align}
with 
\begin{align}
d \tilde \Omega_7^2 = d\beta^2 + \cos^2 \beta d  \Omega^2_3 
+ \sin^2 \beta d \tilde \Omega^2_3 \, 
\end{align}
and $d \tilde \Omega_3$ the metric on $S^3/\ZZ_{N_6}$,
\begin{align}
d \tilde \Omega^2_3 = \frac{1}{4} d\Omega^2_2 + \left[\frac{1}{N_6}
d\psi + \frac{1}{2}(1-\cos \theta ) d\varphi \right]^2 \,.
\end{align}
Here $w$ parametrizes a radial direction and $0 \leq \beta \leq
\frac{\pi}{2}$, $0 \leq \theta \leq \pi$, $0 \leq \varphi, \psi \leq
2\pi$ are angular coordinates.  For massless quarks the factor $F=1$
and the metric is $AdS_4 \times S^7/\ZZ_{N_6}$ with the scale $L=l_p^6
\pi^2 N_2 N_6/2$.

For large $N_2$ this background is a reliable classical geometric
description with a small curvature. As discussed in \cite{Pelc} the
geometry is eleven-dimensional as long as $N_6 \ll N_2^{1/5}$ (or,
equivalently, $\nu \ll N_2^{-4/5}$).  This implies that $\nu$ must be
very small but non-vanishing in the large $N_2$ limit.  As reviewed in
appendix~\ref{appB} of this paper, for $0 \ll \nu \ll 1$ the
appropriate description is given in terms of its ten-dimensional
reduced geometry. For $\nu
\gg 1$ the ten-dimensional curvature becomes large and the geometric
description breaks down. As discussed in detail in \cite{Pelc}, the
field theory becomes weakly coupled at large $\nu$ due to a
simplification of the Feynman diagrams.\footnote{ 
The Feynman diagrams are planar and have only quark
loops, i.e.\ they do not contain loops of adjoint fields.}

Massive quarks are obtained by separating the D2 from the D6-branes by
a distance $m$ as shown on the left hand side of Fig.~\ref{fig1}. This
distance determines the bare quark mass $\frac{m}{2\pi l_s^2}$ in
units of the string tension.  In the eleven-dimensional picture $m$ is
the distance of the M2-branes from the orbifold singularity which are
located at
\begin{align} \label{locationM2}
\beta = \pi/2, \qquad \theta=0, \qquad w=m, \qquad \psi=0
 \textmd{\,\,mod\,\,} 2 \pi \,.
\end{align}
The separation of the branes affects only the function $F$ which is a
measure for the backreaction of the D6-branes. It is given by
\cite{Pelc}
\begin{align} \label{Fm}
   F(w;m)= \frac{1}{N_6} \sum_{k=1}^{N_6} \left(\frac{w_k}{w} \right)^{-3}\,,
  \quad \frac{w_k}{w}=1+\frac{m}{w}-2\sqrt{\frac{m}{w}}
  \sin{\beta} \cos \frac{\theta}{2} \cos\left(\frac{\psi-2\pi k}{N_6}
\right) \,.
\end{align}
\begin{figure} 
\begin{center}
\includegraphics[scale=0.8]{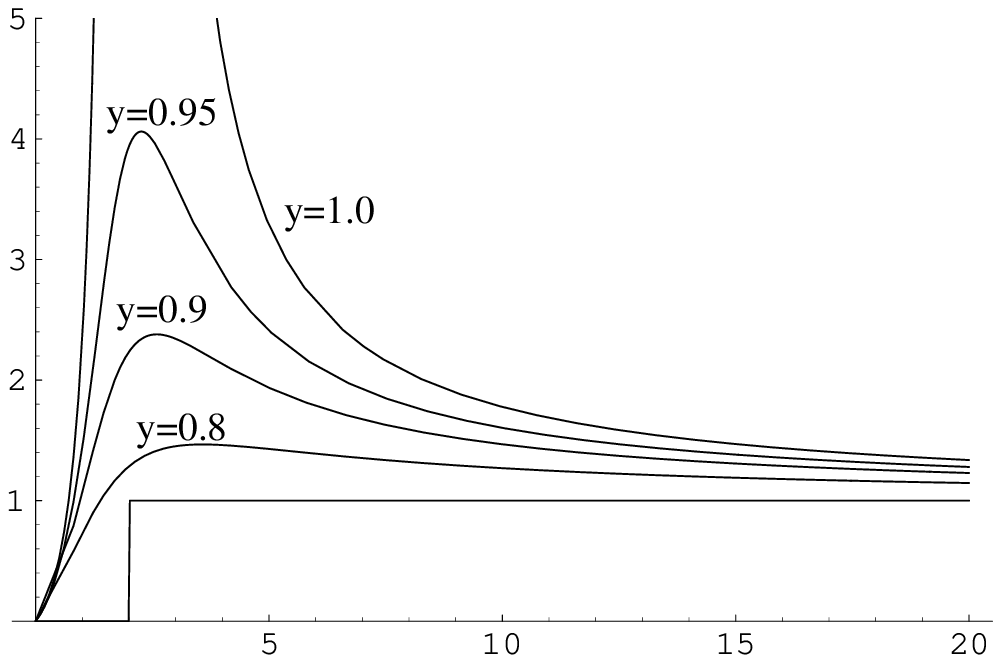}
\end{center} \caption{Plot of $F(w;m=2)$ for different values of
$y=\sin \beta \cos \frac{\theta}{2}$ ($\psi=0$). The step function
$\Theta(w-m)$ is plotted just to guide the eye.} \label{fig2}  
\vspace{-5cm} \hspace{2.3cm} $F(w)$ \vspace{4.5cm} 

\vspace{-2cm} \hspace{12.3cm} $w$ \vspace{1.5cm}
\end{figure}

Plots of $F(w;m)$ are shown in Fig.~\ref{fig2} for different values of
$y=\sin \beta \cos \frac{\theta}{2}$. For fixed angles $\psi$,
$\theta$ and $\beta$, the function $F(w;m)$ typically starts from zero
at $w=0$, increases to a maximum at $w
\approx m$ and decreases to one at $w \rightarrow \infty$.  Asymptotically, 
i.e.~for radii $w \gg m$ the background is $AdS_4 \times
S^7/\ZZ_{N_6}$, which corresponds to the fact that the mass of the
quarks is irrelevant at high energies. The backreaction of the
D6-branes is small at the point where $F$ diverges.\footnote{There is
however no curvature singularity at this point, since $R \sim
1/F^{1/3}$.}  At this point the Pelc-Siebelink background approaches
the near-horizon geometry of the D2-branes which is $AdS_4 \times
S^7$.

\section{Fluctuation-operator map and meson spectrum}

In this section we calculate the lowest-lying meson mass for the D2/D6
configuration. In contrast to calculations in the probe approximation,
the supergravity degrees of freedom dual to meson operators are
already included in the D2/D6 background. We find the meson mass as a function
of the quark mass 
by simply solving the appropriate supergravity wave
equations in the Pelc-Siebelink geometry. Since the 
equations are highly non-linear, the use of numerics is necessary.

This approach is analogous to computations of the glueball mass
spectrum in confining gauge/gravity duals. However, we use a different
field-operator dictionary which maps meson-like operators to
corresponding supergravity fluctuations. In contrast to similar
computations in the probe approximation, we do not need a
Dirac-Born-Infeld action.

\subsection{Fluctuation-operator map} \label{secflucmap}

In the following we discuss the dictionary between supergravity
fluctuations on $AdS_4 \times S^7/\ZZ_{N_6}$ and conformal operators
in the three-dimensional field theory. In particular, we focus on
meson-like operators and their dual supergravity scalars. 

In the field theory we have two kinds of chiral operators: First,
there are composite operators which consist of purely adjoint
fields. An example for such an operator is ${\rm Tr\,}
\phi_{i_1} \hdots \phi_{i_k}$ ($i_j = 1,2,3$) which also occurs in the 
$\N=8$ SYM theory. Secondly, there are meson-like operators which can
be considered as a chain of adjoint fields descending from the $\N=4$
vector multiplet with fundamental matter degrees of freedom (e.g.\
quarks) at both ends. The operator $\bar q X^A_V q$ is an example for
a meson-like operator which is bilinear in fundamental scalars $q$. It is
also possible to construct an operator with the same quantum
numbers as this operator by simply replacing the fundamental
fields by fields of the adjoint hypermultiplet. For example,
the operators
\begin{align}
\bar q X^A_V q {\quad\rm and\quad}  {\rm Tr\,} \bar \phi X^A_V \phi \, \qquad
(\phi = \phi_1, \phi_2)\,
\end{align}
both transform in the $(\bf 3, \bf 1)$ of $SU(2)_V \times SU(2)_H$ and
have the same engineering dimension. This is due to the same scaling
behaviour of the fields in the adjoint and fundamental
hypermultiplets. In contrast, the engineering dimension of fields in
the vector multiplet differs by $1/2$ from that of fields with same
spin in the hypermultiplets.

According to the AdS/CFT prescription, chiral operators with given
quantum numbers belong to the short representations of the
supersymmetry group and couple to corresponding supergravity
fluctuations. The spectrum of supergravity excitations on 
$AdS_4 \times S^7/\ZZ_{N_6}$ is worked out in \cite{Gomis}.
The scalar spectrum is given by the Kaluza-Klein harmonics on $AdS_4
\times S^7$ with mass squared $m^2=\frac{1}{4}k(k-6)$ \mbox{($k \geq
2$)}, upon projecting out degrees of freedom which are not invariant
under the orbifold group. The resulting supergravity spectrum
separates into two classes of fluctuations corresponding to different
representations of the global symmetries: One class is dual to
adjoint operators, while the other class contains fluctuations dual to
meson-like operators. The origin of fluctuations dual to operators
with fundamentals is due to the orbifold projection. This is
remarkable since without projection, all fluctuations are dual to
operators with fields in the adjoint representation only. The orbifold
projection ensures the presence of fluctuations dual to meson
operators.

\begin{figure} 
\begin{center}
\input{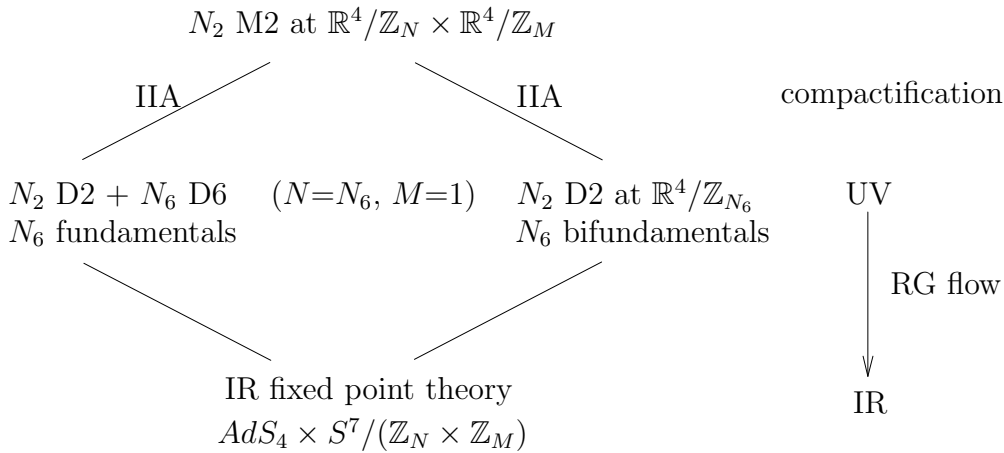}
\end{center} \caption{Mirror symmetry.} \label{mirrorfig}
\end{figure}

One way to understand how the orbifold projection yields fluctuations
dual to meson operators with fields in the fundamental representation
is given by mirror symmetry~\cite{Intriligator:1996ex}. Mirror
symmetry relates two compactifications of M-theory on $\RR^4/\ZZ_N
\times \RR^4/\ZZ_M$ \cite{Ferrara:1998vf}. Considering $N_2$ M2-branes
near the orbifold $\RR^4/\ZZ_N \times \RR^4/\ZZ_M$, type IIA string
theory may be obtained either by compactifying on $\RR^4/\ZZ_N$ or on
$\RR^4/\ZZ_M$. The first compactifaction leads to the intersection of
$N_2$ D2-branes and $N$ D6-branes with the orbifold $\RR^4/\ZZ_M$
placed transverse to the D2-branes but on the world-volume of the
D6-branes. The corresponding world-volume field theory is $SU(N_2)^M$
super Yang-Mills theory with $N$ fundamental hypermultiplets and $M$
bifundamental hypermultiplets.  The second compactification leads to
the same theory with $M$ and $N$ interchanged. In other words, mirror
symmetry interchanges the fundamentals with the bifundamentals. In the
IR both theories flow to the same fixed point theory which is dual to
supergravity on
\mbox{$AdS_4 \times S^7/(\ZZ_N \times \ZZ_M)$}.

We are interested in the case $N=N_6$ and $M=1$ which is shown in
Fig.~\ref{mirrorfig}.  Compactifying a direction transverse to the orbifold,
we obtain $N_2$ D2-branes at the orbifold $\RR^4/\ZZ_{N_6}$. This is the
compactification on the right hand side of Fig.~\ref{mirrorfig}.  The
corresponding field theory of this set-up is a quiver gauge theory with $N_6$
bifundamentals which follows by a projection of the D2-brane theory. On the
gravity side the fluctuation spectrum is obtained by an orbifold projection
from the Kaluza-Klein spectrum on $AdS_4 \times S^7$.  Since on both sides of
the correspondence the spectrum is obtained by an orbifold projection, there
must be some supergravity fluctuations which are
dual to operators involving bifundamentals. Now compactification on the
orbifold circle of $\RR^4/\ZZ_{N_6}$ gives the D2/D6 brane intersection which
is the mirror dual theory shown on the left hand side of Fig.~\ref{mirrorfig}.
Since mirror symmetry maps the bifundamentals to fundamentals in the mirror
dual theory, we expect the existence of fluctuations on $AdS_4 \times
S^7/\ZZ_{N_6}$ which are dual to operators involving fundamentals in the D2/D6
intersection.

Such supergravity fluctuations can indeed be found in the D2/D6
set-up. Let us consider two fluctuations
corresponding to $k=2$ and $k=4$ in the scalar family which are both
tachyons with mass squared $m^2=-2$. These tachyons do not lead to an
instability since their mass lies above the Breitenlohner-Freedman
bound \mbox{$m_{\rm BF}^2=-\frac{9}{4}$}. One of them transforms in
the $(\bf 1, \bf 3)$ of $SU(2)_V \times SU(2)_H$ and is dual to a
$\Delta=1$ operator. The other transforms oppositely in the $(\bf 3,
\bf 1)$ and is dual to a $\Delta=2$ operator.\footnote{The mass of a
p-form on an $AdS_{d+1}$ space is related to the dimension $\Delta$ of
a $(d-p)$-form operator by $ m^2=(\Delta-p)(\Delta + p -d)
\,.$ Here $\Delta=\frac{k}{2}$ and $p=0$.} 
 
The first scalar couples to a dimension $\Delta=1$ superconformal
primary operator in the IR fixed point theory. Chiral operators are in
the short representations of the superconformal algebra and as such,
their dimension is protected from quantum corrections.  Therefore it
is possible to express the operator in terms of fields in the short
distance theory. The chiral primary is the sum of a bilinear in the scalars
$q_f^\alpha=(q_f, \tilde q_f)$ of the fundamental
hypermultiplets ($Q_f, \tilde Q_f$) and a bilinear in the scalars
$\phi^\alpha=(\phi^1, \phi^2)$ of the adjoint hypermultiplet ($\Phi_1,
\Phi_2$). It is given by \cite{Gomis}
\begin{align} \label{operator}
{\cal C}^I=\sum_{f=1}^{N_6}
 \bar q^\alpha_f  \sigma_{\alpha\beta}^I  
q^\beta_f + {\rm Tr\,} \bar \phi^\alpha  \sigma_{\alpha\beta}^I  
\phi^\beta
\qquad (I=1,2,3) \, ,
\end{align}
where $\sigma^I_{\alpha\beta}$ is an element of $SU(2)_H$ such that
the operator transfroms in the $\bf (1, 3)$ of the R-symmetry group
$SU(2)_V \times SU(2)_H$. 

The second scalar is dual to the $\Delta=2$ operator ($A=1,2,3$)
\begin{align}
{\cal O}^A=\sum_{f=1}^{N_6} &\left( \bar \psi^i_f \sigma^A_{ij} \psi^j_f 
+ 2 \bar q_f^\alpha X_V^{Aa} T^a q^\alpha_f \right) 
 \nonumber\\
+ &{\rm Tr\,} \bar \lambda^i \sigma^A_{ij} \lambda^j 
+ {\rm Tr\,}2 \bar \phi^\alpha X_V^{Aa} T^a \phi^\alpha \label{op2}
\end{align}
transforming as $\bf (3,1)$. Here $X^A_V$ are the three scalars
transverse to the D6-branes while $\psi^i = (\psi, \tilde \psi)$ and
$\lambda^i = (\lambda_1, \lambda_2)$ are the spinorial components of
$Q, \tilde Q$ and $\Phi_1, \Phi_2$, respectively.

Both operators ${\cal C}^I$ and ${\cal O}^A$ are unambiguously fixed
by the R-symmetry quantum numbers, the conformal dimension and gauge
invariance. It is not possible to construct an operator out of fields
in the adjoint vector multiplet which has the same quantum
numbers. This can easily be shown for the $\Delta=1$ operator ${\cal
C}^I$. None of the $\Delta=1$ fields in Tab.~\ref{rsym} meets the
required R-symmetry quantum numbers $\bf (1,3)$ such that ${\cal C}^I$
must be a bilinear in the $\Delta=1/2$ scalars $q_f$, $\tilde q_f$,
$q_1$ and $q_2$. The operator ${\cal C}^I$ is the sum of all possible
combinations of these fields. We see that no field of the adjoint
vector multiplet is involved in the construction of ${\cal C}^I$. A
similar argument holds for the $\Delta=2$ operator ${\cal O}^A$.

It is interesting to observe that the supergroup $OSp(4/4)$ is the
same as in the T-dual D3/D5 brane configuration. In the corresponding
defect conformal field theory studied in \cite{DeWolfe, Erdmenger},
the operators ${\cal C}^I$ and ${\cal O}^A$ have a similar structure and are
interpreted as a chiral primary and one of its descendants. Due to
different quantum numbers of the adjoint hypermultiplet in the defect
conformal field theory, the trace terms are absent in the
corresponding defect operators. This is due to the adjoint
hypermultiplet being a four-dimensional bulk field in the defect CFT.
For
instance, the conformal dimension of the three-dimensional scalars $\phi_1$,
$\phi_2$ 
is $\Delta=1/2$, while in four dimensions, scalars have
$\Delta=1$. Another difference to the probe limit 
defect theory is the following: In the D3/D5 configuration the
elements of a series of short representations of $OSp(4/4)$ are given
by the bosonic open string fluctuations on the D5-brane
\cite{DeWolfe}. In our case these elements are provided by the
supergravity fluctuations of the D2/D6 intersection.

\subsection{Scalar meson spectrum}

The derivation of the meson spectrum is very similar to the computation of the
glueball spectrum in confining gauge/gravity dualities. The mass spectrum of
the scalar glueball $J^{PC}=0^{++}$ in pure QCD$_3$, for instance, is obtained
\cite{Csaki, deMelloKoch:1998qs, Zyskin:1998tg, Ooguri:1998hq} by solving
supergravity wave equations in the AdS black hole geometry
\cite{Wittenblackhole}. The computation is based on the duality between the
operator ${\rm Tr}\, F^2$ and the dilaton which satisfy the relation
$m^2=\Delta (\Delta-4)$ of a massless scalar and a dimension four operator. In
order to find the lowest glueball mass modes, one solves the equation of
motion for the dilaton fluctuations by means of the plane wave ansatz
\cite{Csaki}
\begin{align}
\Phi = f(w) e^{ikx} \,,
\end{align}
where $f(w)$ is a function of the radial coordinate $w$ in the black
hole geometry.  The glueball mass $M$ is given by the eigenvalues of
$k^2$ by $M^2=-k^2$. Asymptotically the black hole geometry is AdS,
where the dilaton has the behaviour $\Phi \sim 1, w^{-4}$ and only the
latter is a normalizable solution. For fixed values of $M$, one then
searches for normalizable solutions of the dilaton equation of motion
with asymptotical behaviour~$w^{-4}$.

For the meson spectrum we proceed in the same way using now the
field-operator map derived in the previous section. This is
a map between the operator ${\cal O}^A$ as given by
Eq.~(\ref{op2}) and the supergravity field $\phi$ transforming in the
$(\bf 3, \bf 1)$ of the R-symmetry group. In the supergravity
description we solve the classical equation of motion of the
scalar field $\phi$ with mass squared $m^2=-2$,
\begin{align}
\frac{1}{\sqrt{g}}\partial_\mu \left[\sqrt{g} g^{\mu\nu} 
\partial_\nu \right] \phi + 2 \phi= 0 \,,
\end{align}
on the eleven-dimensional background (\ref{11dPelc}). We assume that
the fluctuations have the plane wave form
\begin{align}
 \phi (w, x) = f(w) \sin k x \,.
\end{align}
The equation of motion for the fluctuations becomes
\begin{align} \label{fluctuationeom}
f''(w) + \frac{4}{w} f'(w) + \left( \frac{16 M^2 F}{w^4} 
+ \frac{2 F^{1/3}}{w^2} \right)f(w) = 0\,,
\end{align}
where we set the scale $L=1$ and defined $'\equiv \partial_w$. Here
$M^2=-k^2$ is the meson mass and $F(w;m)$ is a function of the quark mass
$m$ as defined by Eq.~(\ref{Fm}).  For $w
\rightarrow \infty$ the term including the meson mass can be neglected
and the differential equation reduces to $f''+\frac{4}{w} f' +
\frac{2}{w^2} f =0$. The solutions of (\ref{fluctuationeom}) have the
asymptotic behaviour $w^{-1}$ and $w^{-2}$. This agrees with the
asymptotic behaviour of a supergravity scalar dual to the $\Delta=2$
operator ${\cal O}^A$,
\begin{align}
f(w) \sim c w^{-\Delta} + m w^{\Delta-3} \,.
\end{align}
According to the standard AdS/CFT prescription $m$ is interpreted as a
source of the operator and $c$ as the vev of that operator.

We determine $M^2=-k^2$ in dependence of the quark mass $m$ by the
shooting method. The quark mass enters the equation of motion via the
function $F(w;m)$. For a fixed quark mass
$m$ we numerically integrate the equation of motion
(\ref{fluctuationeom}) with boundary conditions $f \sim w^{-2}$ at
large $w$. Solutions of the equation must be regular at all values of
$w$, such that the allowed $M^2$ solutions can be found by tuning to these
regular forms.

\begin{figure} 
\begin{center}
\includegraphics[scale=1]{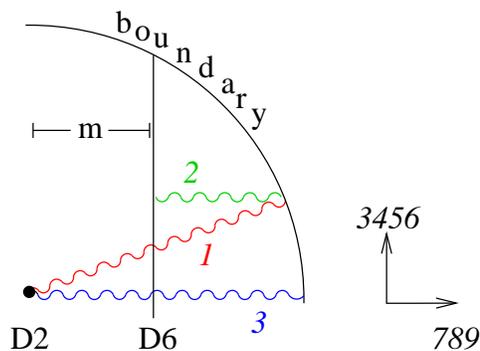}
\end{center} \caption{Self-energies of quarks represented by fundamental 
strings extending from the D2-branes and ending at infinity.}
\label{raddirecfig2}
\end{figure}
Formally the meson mass depends also on the angles $\psi$, $\beta$ and
$\theta$ because of the function~$F$.  Since the metric is not
symmetric in these angles, we face the problem of identifying the
radial direction which corresponds to the energy scale in the field
theory.  Clearly, this is the direction which connects the singularity
at $w=\beta=0$ with the M2-branes. Other directions are not
perpendicular to the M2-branes and cannot be related to an energy
scale in the field theory.

This can be seen by considering the self-energy of a quark which is
represented by a fundamental string stretched between the D2-branes
and the boundary at infinity. Fig.~\ref{raddirecfig2} shows two such
strings, one for $\beta \neq \pi/2$ (string~1), the other for $\beta =
\pi/2$ (string~3). For $\beta \neq \pi/2$ the fundamental string (string 1) is 
perpendicular to the D2-branes but not to the D6-branes.  This string
configuration is not stable and decays to a string perpendicular to
the D6-branes (i.e.\ string~1 decays to string~2 in
Fig.~\ref{raddirecfig2}.) Since now the string ends on the D6-branes
and not on the D2-branes, we cannot identify the string with a charge
in the field theory anymore. The only stable string configuration is
obtained for $\beta = \pi/2$ (string~3). This string is perpendicular
to both the D2 as well as the D6-branes and thus defines the
appropriate radial direction.  For more details on the relation
between the self-energy of quarks and the radial direction in the
geometry see also \cite{Pelc}.

\begin{figure} 
\begin{center}
\includegraphics[scale=1]{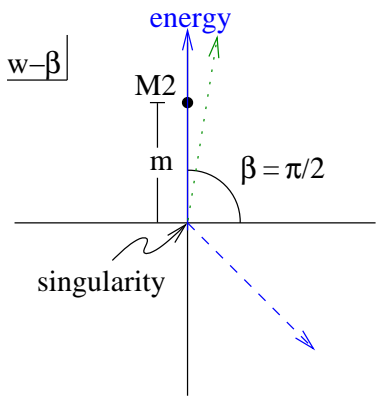} 
\end{center} \caption{The radial direction corresponding to 
the energy scale in the field theory. The dashed direction does not
have an energy interpretation. The dotted arrow represents the
direction which we have chosen for the meson computation and
corresponds to the regularization ($\beta \lesssim \pi/2$).}
\label{raddirecfig}
\end{figure}
We thus choose the values for the angles as in
Eq.~(\ref{locationM2}). This defines a radial direction in the
$w$-$\beta$-plane as shown in Fig.~\ref{raddirecfig}.  Unfortunately,
for $y=\sin(\pi/2) \cos(0)=1$, $F(w;m)$ diverges at $w=m$ which
renders any meson computation impossible. We therefore regularize the
function $F(w;m)$ either by approximating $F(w;m)$ by the step
function $ \Theta(w-m)$ or by deviating from $y=1$. The latter
approximation corresponds to a radial direction with $\beta \lesssim
\pi/2$ which passes close to the M2-branes. 

The Pelc-Siebelink metric possesses a coordinate singularity at the
point $w=m$, $y=1$, where the geometry reduces to an $AdS_4 \times
S^7$ space. This point corresponds to the complete decoupling of
quarks at very low energies. The regularization of $F$ smoothes out
this part of the metric making it suitable for the computation of
meson spectra. This corresponds to a slight modification of the deep
infrared region of the theory which should not affect the qualitative
behaviour of the theory at energies above the quark mass.\footnote{A
more rigorous way would be to find a different coordinate system in
which the metric components are finite in the $AdS_4 \times S^7$
region of the Pelc-Siebelink background.  We leave this for future
investigations.}

\begin{figure} 
\begin{center}
\includegraphics[scale=.8]{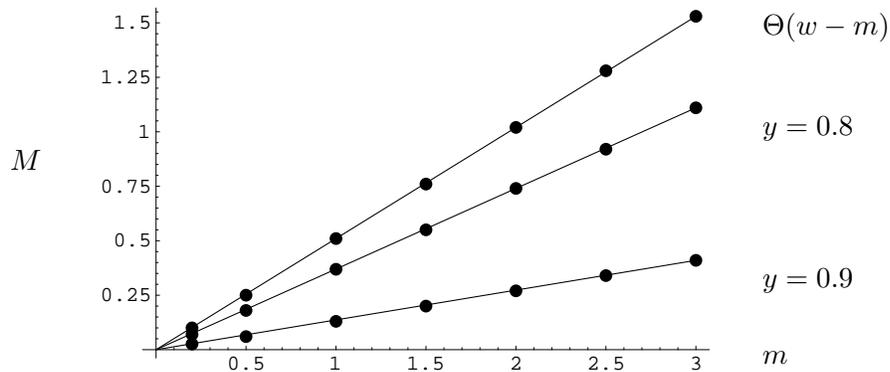}
\end{center}
\caption{Lowest-lying meson mass as a function of the quark mass for three
  different regularizations. The meson mass scales linearly with the quark
  mass. } \label{linear}
\vspace{-6.7cm} \hspace{12.2cm} $\Theta(w-m)$ \vspace{6.2cm}

\vspace{-5.4cm} \hspace{12.2cm} $y=0.8$ \vspace{4.9cm}

\vspace{-3.4cm} \hspace{12.2cm} $y=0.9$ \vspace{2.9cm}

\vspace{-2.4cm} \hspace{12.2cm} $m$ \vspace{1.9cm}

\vspace{-5.0cm} \hspace{2.2cm} $M$ \vspace{4.5cm}
\end{figure}

Fig.~\ref{linear} shows a plot of the meson mass in dependence of the
quark mass for different values of $y$ and for the step function
approximation (labelled by $\Theta$). As in the corresponding
four-dimensional supersymmetric set-up \cite{Kruczenski} we find the
meson mass to depend linearly on the quark mass. This is true for
any of the regularizations we have considered although the slope
of the graphs appears to be strongly regularization dependent.
Nevertheless the linear behaviour is universal.

This behaviour of the meson mass agrees with expectations from field
theory. We may see this by adapting a well-known argument using
effective Lagrangians.  Consider an effective Lagrangian for the
meson-like field $\pi = \bar \psi^i \sigma^3_{ij}
\psi^j+ 2 \bar q^\alpha \sigma q^\alpha$ of the type
\begin{align}
{\cal L} = \frac{1}{2} \left (\partial_\mu \pi \partial^\mu \pi 
- M^2_\pi \pi^2 \right) + ... \quad.
\end{align}
The field $\pi$ is essentially the third component of the operator
${\cal O}^A$ as defined in (\ref{op2}), where we ignored the terms
with fields in the adjoint representation for simplicity.  The meson
mass $M_\pi$ can be expanded in terms of the quark mass $m$ as
\begin{align}
M^2_\pi = m B_0 + m^2 C_0 + ... \quad,
\end{align}
where the coefficient $B_0$ is given by
\begin{align}
B_0 = - \frac{\langle \bar \psi^i \sigma^3_{ij}
\psi^j+ 2 \bar q^\alpha \sigma q^\alpha \rangle}{F^2_{\pi}} \,.
\end{align}
In a supersymmetric theory a quark condensate \cite{Karch} or, in our
case, a vev of the operator ${\cal O}^3$ is forbidden. The reason is
that this operator is an F-Term of the mass operator $m \sigma^3_{ij}
\bar Q^i Q^j$ and a vev for an F-term violates supersymmetry.\footnote{For 
the scalar contribution $\bar q^\alpha \sigma q^\alpha$ recall 
$\langle X_V^3 \rangle \equiv \langle \sigma \rangle = m$ such that
it contributes to $B_0$ rather than~to~$C_0$.}
This argument holds in any dimension.  We conclude that the meson mass
$M$ depends linearly on the quark mass $m$ (at least for small quark
masses). This is different from QCD in which there is no symmetry
constraint to force $B_0=0$ leading to the behaviour $M_\pi \sim
\sqrt{m}$ for small quark masses.

\section{Conclusions}

The results of this paper show that gauge/gravity duality with flavour may be
consistently established beyond the probe approximation.
As an example we have considered
the holographic dual of a three-dimensional $SU(N_c)$
super Yang-Mills theory with $N_f$ fundamental fields in the limit $1
\ll N_f \ll N_c$ with $N_f/N_c$ fixed and finite. We showed
that meson-like operators may consistently be taken to be dual to 
closed strings. In this way we obtain the lowest-lying meson mass
directly from supergravity fluctuations. This is new as compared to previous
examples of gauge/gravity duals with flavour, in which meson operators are
dual to open strings ending on a probe brane.  The meson mass obtained from
supergravity agrees with field theory expectations since it scales linearly
with the quark mass.

It would be interesting to study further examples of 
fundamentals in backgrounds
with backreaction. A similar analysis in four spacetime dimensions
certainly deserves attention, 
perhaps by using the D3/D7 solution
found in \cite{Vaman}. A four-dimensional version of the approach presented
here would also allow for further insight into the holographic version of 
the Witten-Veneziano
mass formula $M^2_{\eta'} \sim \nu \chi_T$ for the $\eta'$ particle (see
\cite{KMB} for a stringy discussion in the probe limit).   
Keeping $\nu=N_f/N_c$ finite and fixed requires the backreaction to be taken
into account. - In a somewhat different direction, it would be interesting to
explore if gauge/gravity duality with flavour may be related to the work of 
\cite{Pons} where the semiclassical
string spectrum for the black hole background dual to pure QCD${}_4$
is discussed.

Another possible extension of this work is to consider mesons in a
three-dimensional non-supersymmetric background.  Ferretti {\em et
al.}\ have shown \cite{Ferretti} that there is a parity preserving
phase of three-dimensional QCD in which the global $U(2N_f)$ symmetry
is broken spontaneously down to $U(N_f) \times U(N_f)$.  The dependence 
of the meson spectrum on the parameter $\nu$ is important: It has
been found in \cite{Wie} that the spontaneous symmetry breaking
scenario does not take place for all values of $\nu=N_f/N_c$. There
are some indications for the existence of a critical value $\nu^{\rm
crit} = 64/3\pi^2 \approx 2.16$ ($N_c$ large) above which the
$U(2N_f)$ flavour symmetry is restored and the theory does not exhibit
dynamical symmetry breaking. It is a challenge to show this
in terms of a dual holographic  set-up.

\bigskip\bigskip

\vspace{0.3cm}

\bigskip

\newpage
\noindent {\bf Acknowledgements}

\medskip

We would like to thank C.\ Nu\~nez for substantial advice at
an early stage of this project. Moreover we are grateful to G.\ Dall'Agata 
and to Z.\ Guralnik for helpful discussions.

Our research was funded by the DFG (Deutsche Forschungsgemeinschaft)
within the Emmy Noether programme, grant ER301/1-3.


\appendix
\section*{Appendix} 

\section{Field theory action} \label{appA}

The field theory corresponding to the low-energy limit of
 the D2/D6 intersection is T-dual to the defect
conformal field theory on the D3/D5 system studied extensively in
\cite{DeWolfe,Erdmenger}. The world-volume theory is $\N=4$ $SU(N_c)$ super 
Yang-Mills theory coupled to $N_c$ adjoint and $N_f$ fundamental
hypermultiplets. We will use a manifest ${\cal N}=2$, $d=3$ superspace
formalism which can be obtained from $\N=1, d=4$ superspace upon
dimensional reduction. Note that only a $U(1)$ subgroup of $SU(2)_V$ is
manifest in $\N=2, d=3$ superspace language.

The field content of this theory is listed in Tab.~\ref{rsym}. Under gauge
transformations the $\N=2$ superfields transform as
\begin{align}
&\Phi_i \rightarrow  e^{-i\Lambda} \Phi_i e^{i\Lambda} \,,\qquad
e^{V} \rightarrow e^{-i \Lambda^{\dagger}} e^{V} e^{i\Lambda}  \,,\\
&Q_f \rightarrow e^{-i\Lambda}Q_f \,, \qquad
\tilde Q_f \rightarrow \tilde Q_f e^{i\Lambda} \, .
\end{align}
where $\Lambda$ is a chiral multiplet. As in $\N=1, d=4$ superspace we
have three chiral and one vector superfield. The fields $Q_f$ and
$\tilde Q_f$ transform in the fundamental and antifundamental
representation of the gauge group, respectively. We can also define a
linear multiplet $\Sigma$ by
\begin{align}
\Sigma \equiv \epsilon_{\alpha\beta} {\bar D}_{\alpha}(e^{-V}D_{\beta}e^{V})
\, \label{nonabelianlinear}
\end{align}
which transforms under gauge transformations as
\begin{align} 
\Sigma\rightarrow e^{-i\Lambda} \Sigma e^{i\Lambda}  \,.
\end{align}

In terms of these superfields the low-energy effective action of the
D2/D6 intersection is given by
\begin{align}
  S = S_{\rm D2} + S_{\rm quarks} \,,
\end{align}
with the ${\N=8}$, $d=3$ super Yang-Mills action 
\begin{align} 
  S_{\rm D2}&= \frac{1}{g^2} \int \! d^3x d^2 \theta d^2
  {\bar\theta}\, {\rm Tr} \left[ \frac{1}{4} \Sigma^2 - 
  e^{-V}{\bar \Phi}_3 e^V \Phi_3 \right] 
  - \sum_{i=1}^2 \int \! d^3x d^2 \theta d^2
  {\bar\theta}\, {\rm Tr}  e^{-V}{\bar \Phi}_i e^V \Phi_i \nonumber\\
&+ \int\! d^3x
  d^2\theta\, {\rm Tr} [\Phi_1, \Phi_2]\Phi_3 + c.c. \label{M3bulk}
\end{align}
and
\begin{align} \label{M3boundary}
S_{\rm quarks} = \int\! d^3x d^2 \theta d^2 {\bar \theta}\,  
({\bar Q_f} e^{V}Q_f + {\tilde Q_f} e^{-V} {\bar {\tilde Q}}_f) 
+ \, \Big( \, \int\! d^3x d^2 \theta \, \tilde Q_f \Phi_3 Q_f   + c.c. \, \Big)
\, . 
\end{align}
The action (\ref{M3boundary}) describes the couplings to the
fundamental hypermultiplets which break supersymmetry from $\N=8$ down
to ${\cal N}=4$. Note that the conformal dimension of the coupling 
constant is $[g]=1/2$. In order to have correct conformal dimensions,
we need the factor $1/g^2$ in front of the kinetic term of the
vector multiplet. 

\section{Pelc-Siebelink background in 10 dimensions} \label{appB}

The ten-dimensional Pelc-Siebelink background is given by \cite{Pelc}
\begin{align} \label{warpedmetric}
 \frac{ds^2}{l_s^2} &= \frac{\sin \beta}{N_6}
 \left[ \frac{w^2}{16 (LF)^{1/2}} dx^2_{||} + (LF)^{1/2} 
\left( \frac{dw^2}{w^2} + d \tilde \Omega_6^2 \right)
 \right] \,, \\ 
 e^\phi&=
 \left(\frac{\pi^2 N_2 F}{2 N_6^5}\right)^{1/4}
 (\sin \beta)^{3/2} \, ,
\end{align}
which asymptotes to a warped product geometry $AdS_4 \times_w X_6$. Here
the $AdS_4$ space is fibered over a compact manifold $X_6$. 

This geometry is a ten-dimensional supergravity solution only for a
certain parameter regime of $\nu$. There are two bounds on $\nu$ given
by requiring both the dilaton as well as the curvature to be
small. The curvature ${\cal R}\sim {\sqrt{\nu}}/{\sin^3 \beta}$ ($\nu
=N_6/N_2$) is small for $\nu \ll 1$. To impede a blow-up of the
dilaton, $N_2$ and $N_6$ must satisfy $N_2^{1/5} \ll N_6$ (or,
equivalently, $N_2^{-4/5} \ll \nu$).  In the large $N_2$ limit the
parameter $\nu$ takes therefore values in the regime $0 \ll \nu \ll
1$. For smaller values of $\nu$ the dilaton becomes very large and a
better description is given by the eleven-dimensional metric~(\ref{11dPelc}).


\end{document}